\documentclass{article}

\usepackage{PRIMEarxiv}

\usepackage[utf8]{inputenc} 
\usepackage{hyperref}       
\usepackage{url}            
\usepackage{booktabs}       
\usepackage{amsfonts}       
\usepackage{nicefrac}       
\usepackage{microtype}      
\usepackage{lipsum}
\usepackage{fancyhdr}       
\usepackage{graphicx}       
\usepackage[labelfont=bf,labelsep=period,singlelinecheck=false]{caption}
\usepackage{amsmath}
\usepackage{multirow}
\usepackage{multicol}
\usepackage{array}
\usepackage[section]{placeins}
\usepackage{tabularx}       
\usepackage{pifont}         
\newcommand{\cmark}{\ding{51}}
\newcommand{\xmark}{\ding{55}}
\graphicspath{{media/}}     

\ifdefined\pdflinkmargin
\else
  \ifdefined\XeTeXLinkMargin
  \fi
\fi

\title{{\upshape Medical Image Analysis with Machine Learning (\textit{MEDiml}): Strategies for Simpler Radiomics}}

\author{
  \textbf{Mahdi Ait Lhaj Loutfi$^1$, Simon Provost$^1$, Élodie Gillard$^2$, Moustafa Amine Bezzahi$^2$, Alexander Chum$^1$, Ben Coull-Neveu$^1$, Juan Duran$^1$, Mariem Kallel$^2$, Hakima Laribi$^2$, Oliver Lefebvre$^2$, Nicolas Longchamps$^2$, Thanh-Tri Nguyen$^1$, Steven Ravary$^1$, Ouael Nedjem Eddine Sahbi$^2$, Jérémy Dana$^{3}$, Martin Vallières$^{1,2*}$}\\
  \vskip 0.2in \center{}   \vskip 0.2in
  $^1$Medical Physics Unit, Department of Oncology, McGill University, Montréal, Canada
  \\$^2$Department of Computer Science, Université de Sherbrooke, Sherbrooke, Canada
  \\$^3$Department of Radiology, Centre de Recherche du Centre Hospitalier de l'Université de Montréal, Montréal, Canada
  \\$^{*}$Corresponding author: Martin Vallières, \texttt{martin.vallieres@mcgill.ca}
}

\begin{document}
\maketitle

\section{Summary}
\label{sec:summary}
Medical images from computed tomography, magnetic resonance, and positron emission tomography contain far more information than a radiologist can extract with the naked eye. Radiomics extracts quantitative features describing shape, texture and intensity patterns within a region of interest (e.g., a tumor) that can then be used to develop machine learning models for predicting clinical outcomes \cite{lambin_radiomics_2012}. 

However, identifying the most predictive features for a given clinical problem remains a challenge, as large feature sets increase the risk of overfitting and reduce the interpretability of the model, particularly in small datasets \cite{issue_small_data_22}. Addressing these challenges often requires programming expertise and familiarity with machine learning methods, limiting the accessibility of radiomics for researchers without specialized technical backgrounds.

To improve the accessibility of radiomics, we present \textit{MEDiml} (\href{https://www.mediml.app}{www.mediml.app}), which stands for Medical Image Analysis with Machine Learning. \textit{MEDiml} is an open-source software available both as a Python package (\textit{MEDiml-pkg}) and as a desktop application (\textit{MEDiml-app}). \textit{MEDiml} covers the full radiomics workflow, including \textbf{image processing}, \textbf{feature extraction}, \textbf{model training} and \textbf{performance evaluation}.


\textit{MEDiml} is built around complexity minimization \cite{loutfi_complexity_2024}. Features are grouped into complexity levels according to how much computation they require, from shape and size, to intensity, to texture, to features from filtered images. Simpler models are less prone to overfitting in small datasets, are more generalizable, and are easier for clinicians to interpret \cite{overf_radiomics}. For this purpose, \textit{MEDiml} trains a model at each level and compares the levels with statistical tests, such that the simplest level with sufficient predictive performance is retained. Moreover, feature extraction in \textit{MEDiml} follows international standards for radiomics extraction, set by the Image Biomarker Standardization Initiative (IBSI) guidelines \cite{ibsi_1, ibsi_2}, making feature extraction reproducible across studies.

Finally, clinical researchers without programming experience can design and execute radiomics experiments through a graphical interface, while other users can directly interact with the Python package or export interface-generated workflows as executable Python code.

\section{Statement of Need}
\label{sec:statement_of_need}
Radiomics has a well-documented reproducibility problem, as feature values can shift substantially depending on image preprocessing choices, discretization parameters, and whether the software follows standardized definitions. A study by Zwanenburg \textit{et al.} \cite{zwanenburg_assessing_2019} found that only a subset of radiomic features remained robust under image perturbations such as noise addition, rotation and contour variation. Although clear guidelines and standards now exist, the IBSI being the most significant, clinical translation remains poor \cite{how_to_7, vulnerabilities_8}.

The high dimensionality of radiomic feature sets is another significant challenge. Feature sets can exceed 10,000 values per scan, especially when image filters are applied \cite{radiomics_10k_6}. However, medical imaging datasets are rarely large enough to handle the required ratio of features to patients needed to prevent overfitting in specialized predictive modeling tasks \cite{issue_small_data_22}. Furthermore, the predictive value of radiomic features is highly dependent on the clinical context. For example, in lung cancer survival prediction, tumor volume alone has been shown to carry most of the prognostic signal, making texture features redundant \cite{vulnerabilities_8}, whereas wavelet-transformed texture features were able to distinguish tissue subtypes in other applications \cite{depeursinge_lung_9}.

Finally, the successful application of radiomics requires bridging the expertise of interdisciplinary teams: clinical researchers who understand the medical context, and computer scientists who can implement the analytical workflows. Nonetheless, tools designed primarily for one group may not meet the needs of the other.


\section{State of the Field}
\label{sec:state_of_field}
Several radiomics tools exist, ranging from open-source feature extraction libraries to graphical platforms. Their coverage of feature extraction, graphical interaction and predictive modeling varies, as does their reported coverage of the IBSI recommendations (see Table~\ref{tab:comparison} for a comparison with \textit{MEDiml}).

\begin{table}[ht]
\centering
\footnotesize
\renewcommand{\arraystretch}{1.3}
\caption{\textbf{Comparison of \textit{MEDiml} to Commonly Used Alternative Radiomics Extraction Tools.} GUI: graphical user interface. A check mark indicates that a capability is available; a cross indicates that it is unavailable. ``Partial'' denotes incomplete compliance with the IBSI, and N/A means not applicable.}\medskip
\begin{tabularx}{\textwidth}{|>{\raggedright\arraybackslash}p{0.13\textwidth}|>{\raggedright\arraybackslash}X|>{\raggedright\arraybackslash}X|>{\raggedright\arraybackslash}X|*{5}{>{\centering\arraybackslash}X|}}
\hline
\textbf{Package/Software} & \textbf{Type} & \textbf{Language} & \textbf{IBSI compliance} \cite{ibsi_1, ibsi_2} & \textbf{Own radiomics engine} & \textbf{Machine learning integration} & \textbf{Local execution} & \textbf{Code generation} & \textbf{Open-source} \\ \hline
PyRadiomics \cite{vangriethuysen_pyradiomics_2017} & Library                         & Python             & Partial         & \cmark & \xmark  & \cmark & N/A    & \cmark \\ \hline
3D Slicer (SlicerRadiomics) \cite{fedorov_slicer_2012, vangriethuysen_pyradiomics_2017} & GUI (Plugin)                    & Python             & Via PyRadiomics & \xmark & \xmark  & \cmark & \xmark & \cmark \\ \hline
QuantImage v2 \cite{quantimage_v2} & GUI (web)                       & Python, JavaScript & Via PyRadiomics & \xmark & \cmark  & \xmark & \xmark & \cmark \\ \hline
MIRP \cite{zwanenburg_mirp_2024} & Library                         & Python             & IBSI 1 \& 2     & \cmark & \xmark  & \cmark & N/A    & \cmark \\ \hline
LIFEx \cite{nioche_lifex_2018} & GUI (Freeware)                  & Java               & Partial         & \xmark & \cmark  & \cmark & \xmark & Free / Closed Source \\ \hline
CERR \cite{cerr_20} & Library and GUI                 & MATLAB             & IBSI 1 \& 2     & \cmark & Partial & \cmark & \xmark & \cmark \\ \hline
MITK \cite{wolf_mitk_2005} & Library and GUI (desktop app)   & C++                & Partial         & \cmark & Partial & \cmark & \xmark & \cmark \\ \hline
PySERA \cite{salmanpour_pysera_2026} & Library                         & Python             & IBSI 1 \& 2     & \cmark & \cmark  & \cmark & N/A    & \cmark \\ \hline
\textbf{\textit{MEDiml}} \cite{loutfi_complexity_2024} & \textbf{Library and GUI (desktop app)} & \textbf{Python, JavaScript} & \textbf{IBSI 1 \& 2} & \cmark & \cmark & \cmark & \cmark & \cmark \\ \hline
\end{tabularx}
\label{tab:comparison}
\end{table}

\textit{MEDiml}'s specific contribution is the combination of IBSI compliance, a zero-code graphical interface, a Python package, integrated machine learning tools, and a code-generation feature that connects the graphical and programmatic options. In many of the reviewed tools, predictive modeling is either absent or limited to isolated functionalities, requiring users to export features to external software for feature selection, model development, or statistical analysis. Moreover, none of these tools connects graphical and programmatic interfaces through code generation.

\section{Software Design}
\label{sec:software_design}

\begin{figure}[!htbp]
      \centering
    \includegraphics[scale=0.180]{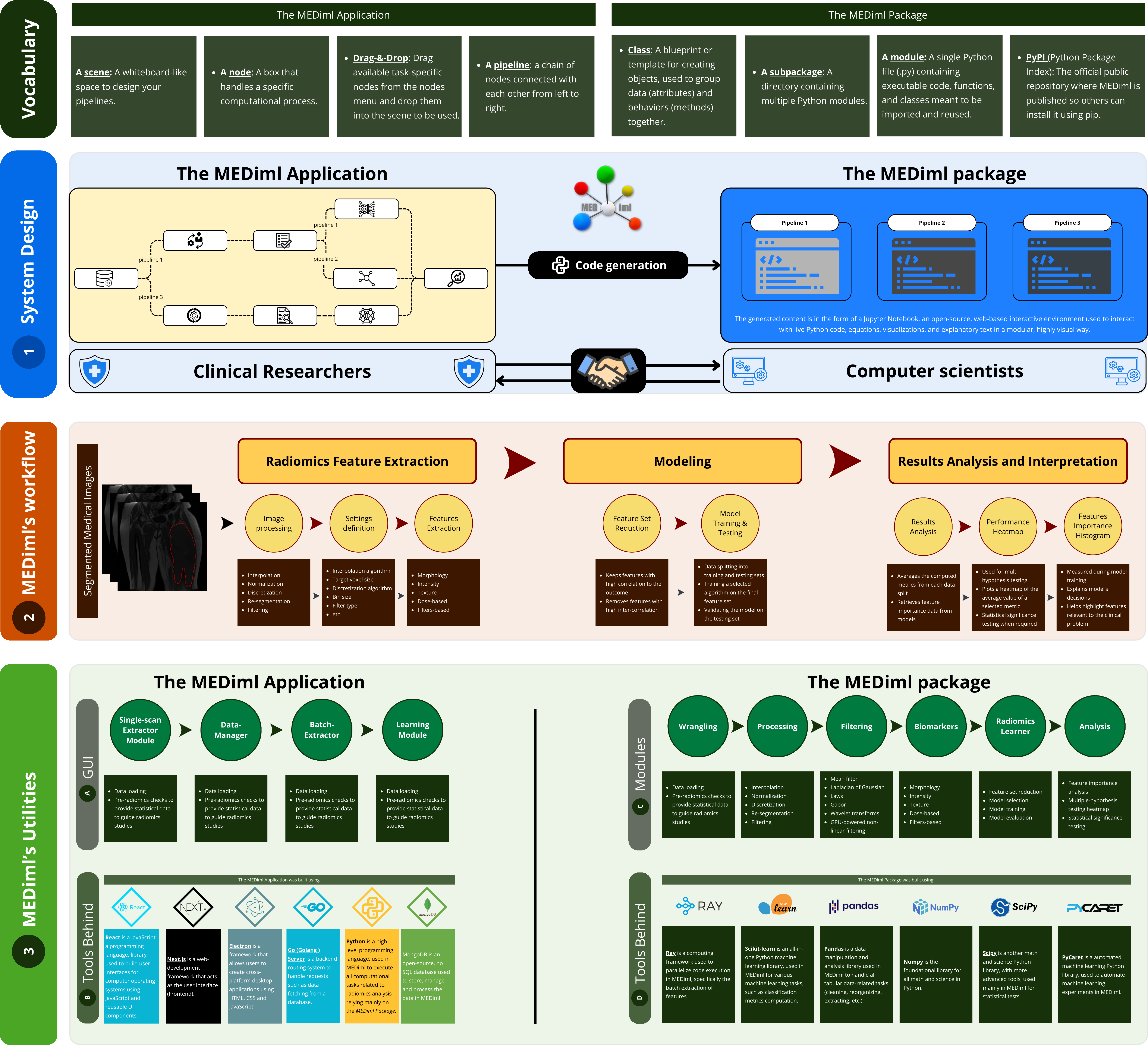}
    \caption{\textbf{Overview of the \textit{MEDiml} Application and Python Package.} Block 1 shows the system design of both components and the code-generation feature that connects graphical workflows with Python code. Block 2 presents the analysis workflow, which starts with feature extraction from processed medical images, and continues with feature set reduction, model training, and results analysis. Block 3 shows the modules available through each component and the technologies on which they depend.}
    \label{fig:workflow}
\end{figure}

A key architectural characteristic of \textit{MEDiml} is that the graphical interface and the Python package run on the same underlying computation, rather than being independent products, to ensure methodological consistency and reproducibility across components.

The default radiomic workflow in \textit{MEDiml} includes the following steps: feature extraction, feature cleaning to remove constant, quasi-constant, and duplicate features, as well as feature normalization using ComBat \cite{combat} to mitigate non-biological batch effects arising from differences in scanner manufacturers, acquisition protocols, or reconstruction settings. Feature selection is then performed using the False Discovery Avoidance (FDA) approach \cite{fda_21} to identify the smallest subset of non-redundant predictive features associated to the clinical outcome. The selected features are subsequently used to train machine learning models using the \textit{PyCaret} framework \cite{ali_pycaret_2024}, with an integrated analysis mode that explains predictive decisions using individual feature importance driving the model's performance. This workflow is depicted in Figure \ref{fig:workflow}, Block 2.

\begin{figure}[!htbp]
      \centering
    \includegraphics[scale=0.380]{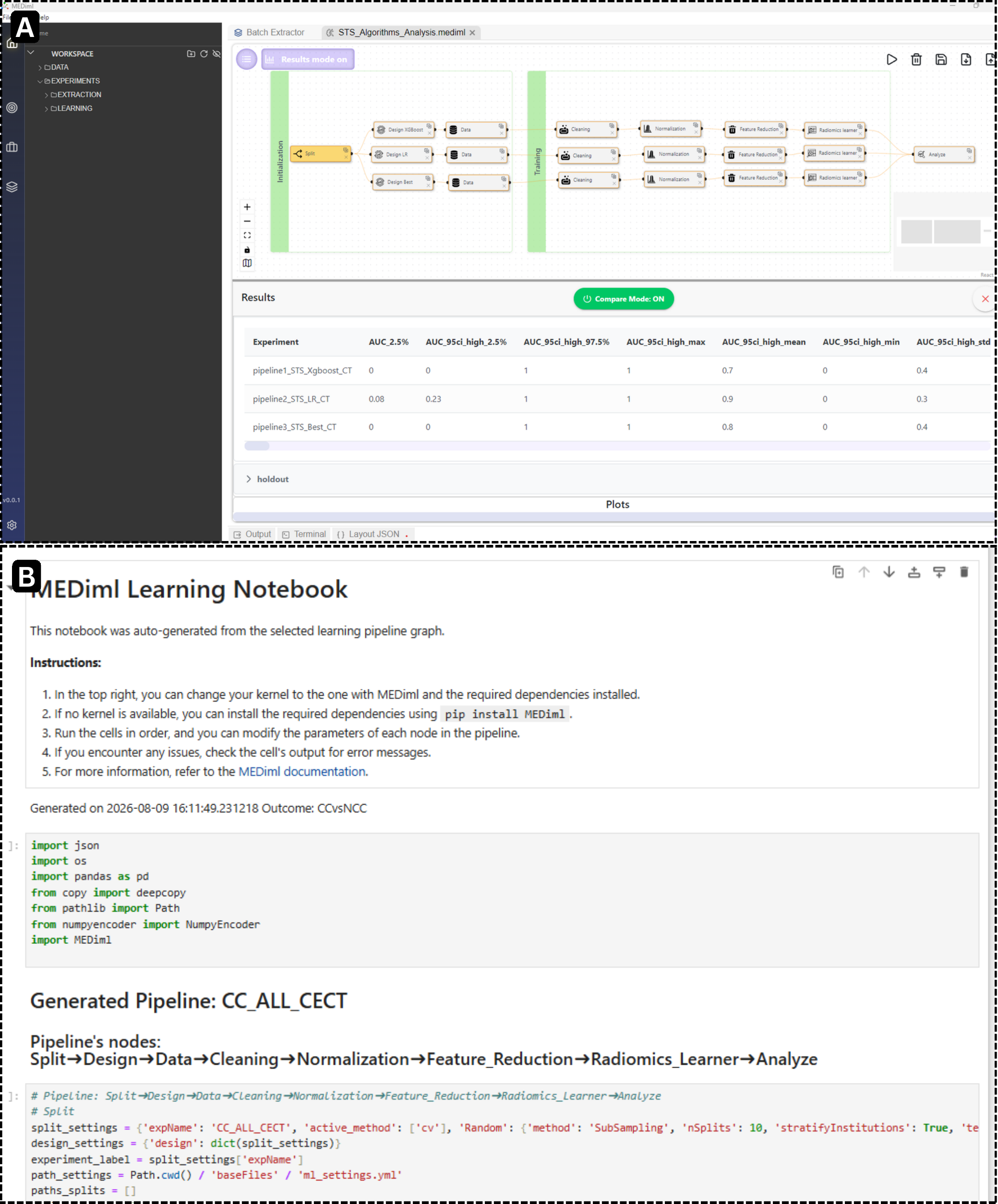}
    \caption{\textbf{A. The \textit{MEDiml} Graphical Interface, Showing the Learning Module}. This module provides a visual workspace for configuring, conducting and analyzing machine learning experiments. \textbf{B. Example of a Generated Jupyter Notebook \cite{kluyver_jupyter_2016}}. The notebook is exported using the code generation feature and contains executable Python code corresponding to a graphically configured workflow.}
    \label{fig:mediml_depiction}
\end{figure}

The \textit{MEDiml} Python package (\textit{MEDiml-pkg}) provides the computational framework used by both programmatic analyses and the graphical application. It is compliant with IBSI 1 and IBSI 2 \cite{ibsi_1, ibsi_2}, builds upon widely adopted scientific Python libraries, including \textit{NumPy} \cite{harris_numpy_2020}, \textit{Pandas} \cite{pandas_2020} and \textit{PyCaret} \cite{ali_pycaret_2024}, and uses a modular architecture where each stage of the workflow is implemented as an independent component (Figure \ref{fig:workflow}, Block 3, provides a summary of the architecture):

\begin{itemize}
    \item \textbf{\textit{Wrangling}}: This module covers data loading through the \textit{DataManager} class, where all imaging scans are loaded, along with their corresponding masks, and prepared for subsequent processing. This module also provides users with statistics, including voxel intensity distribution analyses, to assist in the selection of the appropriate radiomic feature extraction parameters.
    \item \textbf{\textit{Processing}}: This module implements all methods required for data processing steps recommended by the IBSI, including interpolation, region of interest (ROI) extraction, re-segmentation, intensity discretization, and Standardized Uptake Value (SUV) computation for Positron Emission Tomography (PET) imaging, which are fundamental to radiomics analyses \cite{ibsi_1}.
    \item \textbf{\textit{Filters}}: This optional module applies filtering techniques on images prior to feature extraction, to reduce noise, highlight anatomical structures, and generate new image representations, either using linear filters (mean, Laplacian-of-Gaussian, Laws, Gabor and wavelets) or nonlinear textural filters based on the gray-level co-occurrence matrix (GLCM) \cite{sebastian_glcm_2012}, implemented in the \textit{TexturalFilter} class. The latter are optimized to run on a Graphics Processing Unit (GPU) for computational efficiency \cite{buck_cuda_2007, klockner_pycuda_2012}.
    \item \textbf{\textit{Biomarkers}}: This core module performs the extraction of all IBSI-compliant radiomic feature families, from original or filtered images, for individual scans or entire cohorts through the \textit{BatchExtractor} class. The \textit{Ray} library \cite{moritz_ray_2018} is used to parallelize extraction tasks for faster computation.
    \item \textbf{\textit{Learning}}: This module encompasses the complete machine learning workflow. The \textit{Feature Set Reduction (FSR)} class reduces the feature sets using the FDA method \cite{fda_21}, and the \textit{RadiomicsLearner} class trains and evaluates models across radiomic complexity levels, using any \textit{PyCaret} algorithm with $k$-fold cross-validation \cite{arlot_cv_2010}, random subsampling or bootstrapping. The \textit{Results} class then compares complexity levels using statistical significance tests (\textit{DeLong's} \cite{delong_1988, delong_24}, \textit{Wilcoxon} \cite{wilcoxon_1945}, \textit{paired} \cite{student_ttest_1908} and \textit{Nadeau-Bengio corrected t-tests} \cite{nadeau_bengio_2003}) and performance heatmaps, and interprets models through feature-importance histograms and Explanation Trees, which trace predictive contributions from the complexity level to filters, feature families and individual features. Finally, the optimal pipeline is retrained on the entire learning set, evaluated on a holdout set if provided, and exported for deployment.
    \item \textbf{\textit{Utils}}: This module contains general utility and shared components that support the operation of the other software modules.
\end{itemize}

The \textit{MEDiml} desktop application (\textit{MEDiml-app}) is a zero-code graphical interface in which every action invokes the same underlying Python package methods (Figure \ref{fig:mediml_depiction}A). The application is built using Electron \cite{electron}, which packages a web-based interface as a native desktop application. Its drag-and-drop, node-based workflow editor is implemented using React Flow \cite{reactflow} (\url{https://reactflow.dev}), a package for constructing graph-based editors, and is organized around three hierarchical concepts. A \textbf{node} represents a specific processing step, such as data loading, feature cleaning, or model training, with configurable parameters and well-defined inputs and outputs. A \textbf{pipeline} is an ordered sequence of connected nodes, making the structure of the experiment explicit. A \textbf{scene} contains one or more pipelines that can be executed and compared simultaneously, and is the unit for saving, sharing and reproducing analyses.

The application exposes the package through four modules that follow the stages of a radiomics study:

\begin{itemize}
    \item \textbf{\textit{Single-Scan Extraction}}: extracts features from an individual image and its mask to explore and validate processing and extraction settings.
    \item \textbf{\textit{DataManager}}: provides graphical access to the \textit{DataManager} class to register a dataset and run exploratory analyses that guide the choice of extraction settings.
    \item \textbf{\textit{BatchExtractor}}: runs parallelized extraction across the full cohort, with settings configured per imaging modality, and produces a single feature table that can be split by feature family (morphological, intensity and texture) through its feature categorization step.
    \item \textbf{\textit{Learning}}: provides nodes for each stage of predictive modeling (Split, Design, Data, Cleaning, Normalization, Feature Reduction, Radiomics Learner and Analyze), allowing users to train, compare and interpret models across radiomic complexity levels, datasets or imaging modalities within the same scene.
\end{itemize}

The full \textit{MEDiml} architecture is thoroughly explained in a published reference manual \cite{mediml_ref_manual}.

\subsection{Code Generation}
\label{sec:code_gen}
Since both interfaces rely on the same computational engine, graphical and programmatic workflows are functionally equivalent: a graphically configured \textbf{pipeline}, including its feature extraction settings, machine learning models and cross-validation parameters, is executed by the Python package (see Figure \ref{fig:workflow}, Block 1). Users can therefore export any \textbf{pipeline} directly from the interface as executable Python code in a Jupyter Notebook \cite{kluyver_jupyter_2016} format, with each \textbf{node} and its configuration converted into code (see Figure \ref{fig:mediml_depiction}B).

\section{Research Impact Statement}
\label{sec:impact}
\textit{MEDiml} has been applied in two radiomics studies \cite{loutfi_complexity_2024, teodora}, in which the software handled image processing, feature extraction, predictive modeling, and analyses of results. In the first study, \textit{MEDiml} was applied to five cancer imaging datasets, totaling 2,104 scans; the software successfully identified, for each task, the simplest complexity level achieving optimal predictive performance, focusing on less complex models. The associated results are presented in Table \ref{tab:results}.

\begin{table}[ht]
\centering
\footnotesize
\renewcommand{\arraystretch}{1.3}
\caption{\textbf{\textit{MEDiml}'s Radiomics Analysis Results.} AUC: area under the receiver operating characteristic curve \cite{fawcett_roc_2006}; IDH1: isocitrate dehydrogenase 1; CECT: contrast-enhanced computed tomography; MRI: magnetic resonance imaging.}\medskip
\begin{tabularx}{\textwidth}{|>{\raggedright\arraybackslash}X|>{\raggedright\arraybackslash}X|>{\centering\arraybackslash}p{0.20\textwidth}|>{\centering\arraybackslash}p{0.09\textwidth}|>{\centering\arraybackslash}p{0.09\textwidth}|>{\centering\arraybackslash}p{0.15\textwidth}|}
\hline
\textbf{Cohort} & \textbf{Clinical Endpoint} & \textbf{AUC \newline [Confidence Interval]} & \textbf{Sensitivity} & \textbf{Specificity} & \textbf{Optimal \newline Feature Type} \\ \hline
Non-small cell lung cancer (NSCLC) & Histological subtype & 0.76\par\mbox{[0.71, 0.80]} & 0.73 & 0.71 & Intensity features \\ \hline
Low-grade glioma & IDH1 mutation & 0.68\par\mbox{[0.60, 0.75]} & 0.68 & 0.69 & Morphological features \\ \hline
Meningioma & Pathological tumor grade & 0.65\par\mbox{[0.59, 0.72]} & 0.64 & 0.62 & Morphological features \\ \hline
Renal cell carcinoma (RCC) - CECT & Subtype discrimination & 0.82\par\mbox{[0.76, 0.88]} & 0.77 & 0.78 & Intensity features \\ \hline
Renal cell carcinoma - MRI & Subtype discrimination & 0.72\par\mbox{[0.68, 0.77]} & 0.71 & 0.65 & Texture features \\ \hline
\end{tabularx}
\label{tab:results}
\end{table}

\begin{figure}[!htbp]
  \centering
  \includegraphics[scale=0.6]{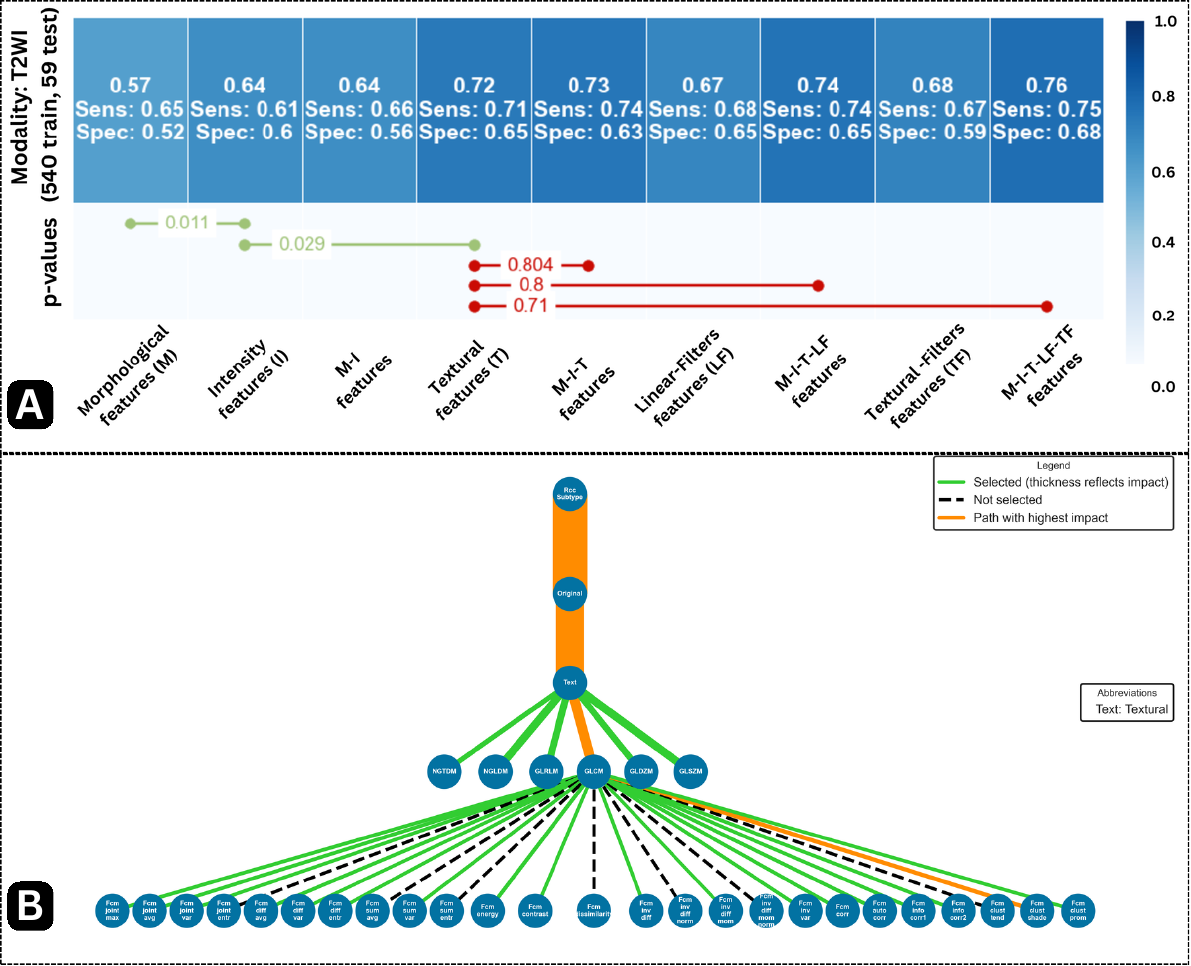}
  \caption{\textbf{Complexity-level Analysis of Renal Cell Carcinoma (RCC) Subtype Classification (clear cell vs. papillary) From T2-weighted MRI.} \textbf{A.} Mean AUC, sensitivity (\textit{Sens}) and specificity (\textit{Spec}) across test folds for each complexity level, ordered from simplest (left) to most complex (right). Bottom rows show \textit{Delong}'s p-values between levels: green, significant improvement (p < 0.05); red, no significant improvement. Textural features (T) were selected as the optimal level, as no more complex level performed significantly better. \textbf{B.} Explanation Tree of the selected model, tracing feature importance from the original (unfiltered) image to textural feature families and individual GLCM features. Line thickness reflects importance, dashed lines mark unselected features, and the orange path leads to the most important feature (GLCM cluster tendency).}
  \label{fig:importance_tree}
\end{figure}

These studies demonstrated the utility of \textit{MEDiml}'s complexity-minimization framework, emphasizing simpler and interpretable models. For example, in MRI-based RCC subtype classification, texture features were selected as the optimal complexity level (Figure \ref{fig:importance_tree}A), and the Explanation Tree traced their contribution down to individual features (Figure \ref{fig:importance_tree}B).

\textit{MEDiml} started as a tool primarily to support the core developer's own research, but it has since expanded to serve interdisciplinary teams. The package has been downloaded over 500 times from the Python Package Index (PyPI) to this date, reflecting a growing user base beyond its origins.

To support users, all information about the software is available on the official website: \url{https://mediml.app}. Overall, \textit{MEDiml} provides the following set of resources:

\begin{enumerate}
    \item \textbf{For the Python package:}
    \begin{enumerate}
        \item Package: \href{https://pypi.org/project/mediml/}{PyPI}.
        \item Documentation: \href{https://mediml.readthedocs.io/en/latest/}{ReadTheDocs}.
        \item Tutorials: \href{https://github.com/MEDomicsLab/MEDiml-pkg/tree/dev/notebooks}{Jupyter Notebooks}.
        \item Source Code: \href{https://github.com/MEDomicsLab/MEDiml-pkg}{GitHub}.
    \end{enumerate}
    \item \textbf{For the desktop application:}
    \begin{enumerate}
        \item Application: \href{https://www.mediml.app/\#download}{Download Page}.
        \item Documentation on \href{https://medomicslab.gitbook.io/mediml-app-docs}{GitBook}.
        \item Video tutorials: \href{https://www.youtube.com/playlist?list=PLEPy2VhC4-D5Eg-UxRyTtmUZRh-D5m_Ru}{YouTube}.
        \item Source Code: \href{https://github.com/MEDomicsLab/MEDiml-app/}{GitHub}.
    \end{enumerate}
    \item \textbf{A community \href{https://discord.gg/GjMW72fE9T}{Discord server}.}
\end{enumerate}

\textit{MEDiml} is maintained and actively improved by the MEDomicsLab research group (\url{https://www.medomicslab.com}) at McGill University.

\section*{Acknowledgments}
We thank Guillaume Blain, Sarah Denis, and Andréanne Allaire for their valuable contributions to the development and enhancement of \textit{MEDiml}'s user interface. We also thank Corentin Gauthier, Clarisse Cheng, and Anthonin Falk for their valuable contributions to the development of the \textit{MEDiml} package. Finally, we gratefully acknowledge Martin Lepage for co-supervising a significant portion of this work \cite{loutfi_complexity_2024}.

Martin Vallières acknowledges the support of Unité de soutien au Système de Santé Apprenant (SSA) Québec.

\bibliographystyle{unsrturl}
\bibliography{references}

\end{document}